\documentclass[aps,prl,letterpaper,amsmath,amssymb,superscriptaddress,nofootinbib,twocolumn]{revtex4-2}

\usepackage{CJK}
\usepackage{graphicx}
\usepackage[usenames]{color}
\usepackage[colorlinks=true,linkcolor=magenta,citecolor=magenta,anchorcolor=green,urlcolor=magenta]{hyperref}
\usepackage{mathrsfs}
\usepackage[percent]{overpic}
\usepackage{tabularx}
\usepackage{multirow}
\usepackage{orcidlink}
\usepackage{braket}

\makeatletter

\newcommand{\Rmnum}[1]{\expandafter\@slowromancap\romannumeral #1@}
\makeatother

\def\be{\begin{equation}}
\def\ee{\end{equation}}
\def\ba{\begin{aligned}}
\def\ea{\end{aligned}}
\def\bea{\begin{eqnarray}}
\def\eea{\end{eqnarray}}

\begin{document}
\begin{CJK*}{UTF8}{}
\title{Numerical renormalization group integrated Hamiltonian truncation: Toward
generic deformation of integrable lattice models}
\author{Xiaodong He \CJKfamily{gbsn}(何晓东)~\orcidlink{0009-0008-4724-9483}}
\affiliation{Tsung-Dao Lee Institute,
Shanghai Jiao Tong University, Shanghai, 201210, China}

\author{Xiao Wang \CJKfamily{gbsn}(王骁)~\orcidlink{0000-0003-2898-3355}}
\affiliation{Department of Physics, Cornell University, Ithaca, NY, USA}

\author{Jianda Wu \CJKfamily{gbsn}(吴建达)~\orcidlink{0000-0002-3571-3348}}
\altaffiliation{wujd@sjtu.edu.cn}
\affiliation{Tsung-Dao Lee Institute,
Shanghai Jiao Tong University, Shanghai, 201210, China}
\affiliation{School of Physics \& Astronomy, Shanghai Jiao Tong University, Shanghai, 200240, China}
\affiliation{Shanghai Branch, Hefei National Laboratory, Shanghai 201315, China}

\date{\today}
\begin{abstract}
We present a hybrid lattice Hamiltonian truncation method that integrates the numerical renormalization group (NRG) 
with a truncated lattice integrable spectrum. The technique is tailored for generic deformations of integrable lattice models, 
where the NRG enables a controlled incorporation of high-energy states. 
The method extends the basis set more effectively and efficiently than brute-force truncation, meanwhile significantly reducing errors.
We show its capability on two paradigmatic models: an Ising chain in a magnetic field and a quantum Ising ladder. 
The resulting dynamical structure factors accurately capture the essential low-energy physics, including 
the $E_8$ and $\mathcal{D}_8^{(1)}$ excitations of the former and later models, respectively,
demonstrating the approach's computational efficiency and high performance.
\end{abstract}

\maketitle
\end{CJK*}

\textit{Introduction.}--- 
The study of quantum many-body systems relies on a diverse set of analytical and numerical techniques, 
including Bethe ansatz~\cite{Bethe, RevModPhys.85.1633,PhysRevB.96.220401,10.1143PTP.48.2187},
Laughlin wave function~\cite{PhysRevLett.51.605,PhysRevLett.50.1395}, 
numerical renormalization group (NRG)~\cite{RevModPhys.80.395,PhysRevB.64.045103,PhysRevB.74.245114,PhysRevLett.99.076402}, 
density matrix renormalization group (DMRG)~\cite{PhysRevLett.69.2863,SCHOLLWOCK201196,RevModPhys.93.045003,Orus:2013kga}, quantum Monte Carlo~\cite{RevModPhys.73.33, PhysRevD.24.2278, PhysRevLett.94.170201,RevModPhys.94.015006}, and quantum neural states~\cite{PhysRevLett.118.216401,PhysRevB.97.085104,PRXQuantum.2.040201}
etc. 
Among these, the truncated spectrum approach (TSA)~\cite{Yurov:1989yu, Yurov:1991my} 
has shown advances in studying low-energy of quantum field theory, 
proving its effectiveness from perturbative 
to non-perturbative regimes~\cite{fitzpatrick2022snowmasswhitepaperhamiltonian, PhysRevD.91.025005, EliasMiro:2021aof, Yurov:1990kv, Lassig:1990xy, Katz:2016hxp, Katz:2014uoa, Anand:2021qnd, Hogervorst:2018otc, Hogervorst:2021spa, Chen:2021bmm, Katz:2013qua, Fitzpatrick:2018ttk, Chen:2022zms, Fitzpatrick:2023aqm, Chen:2021pgx, PhysRevD.102.065001, Anand:2020qnp, Rakovszky:2016ugs, Hodsagi:2018sul}.
Konik and Adamov further advanced the TSA by integrating the NRG procedure, 
a development that substantially reduces the dimension of the truncated spectrum while
also improving its accuracy~\cite{PhysRevLett.98.147205, Brandino:2010sv}.
In recent years, the TSA has been further extended to various 
lattice quantum many-body models and applied to the study of strongly correlated quantum systems~\cite{Albert:2024pwv}. 
For instance, truncated Landau levels have been utilized to detect 
higher-dimensional quantum criticality~\cite{PhysRevX.13.021009,Hu2024}. 
Remarkably, the truncated string state space approach, which hybridizes string states truncation with spectrum truncation, 
provides excellent access to the low-energy spectrum and thermodynamics of the non-integrable spin-1/2 Heisenberg chain~\cite{PhysRevB.108.L020402, PhysRevB.109.214421}. 

The working subspace in these lattice extensions of TSA is much smaller than the full Hilbert space, 
allowing calculations at system sizes beyond exact diagonalization (ED) and providing access to low-lying eigenstates that are not easily obtained by DMRG.
Despite these advantages, the TSA still suffers from limited control over the truncation dimension, which restricts its applicability to complex lattice Hamiltonians.
For example, in a weakly coupled critical Ising ladder, the ladder structure yields a coupled (tensor-product) Hilbert space, which causes the total number of basis states required for Hamiltonian truncation to grow rapidly and further renders standard truncated-Hamiltonian methods ineffective.

To address these limitations, we build upon the truncated lattice integrable spectrum approach (TLISA)~\cite{rw2m-33v6}, which extends the TSA to lattice models by utilizing integrability to construct a truncated Hilbert space focusing on low-energy physics, and then devise a calculation protocol 
which hybrids the NRG and TLISA, dubbed as NRG-TLISA. 
It constructs a low-energy subspace and refines it via iteratively and systematically incorporating
high-energy sectors' contribution, 
arriving at a renormalization of low-energy sector without brute-force cutoffs.
The framework yields renormalized truncated eigen wavefunctions 
with manageable Hilbert space dimensions, enabling detailed investigation of many-body 
dynamics and non-equilibrium physics for generic deformation
of integrable lattice models.
By accessing excited states, the method provides a promising 
route to calculate dynamical structure factors (DSFs) with considerable system sizes.
We demonstrate the NRG-TLISA on a quantum Ising chain with magnetic field 
and a quantum Ising ladder with both open and periodic boundary conditions. 
The DSFs from the NRG-TLISA
show excellent agreement with the analytical
results from the two models at their integrable points.

\textit{Methods.}--- For a lattice system of size $\mathcal V$, the Hamiltonian can be divided into two parts
\begin{equation}
    \mathcal{H} = \mathcal{H}_{0} + \sum_{i} \mathcal{O}_{i},
    \label{Eq:HNRG}
\end{equation}
where $\mathcal{H}_0$ denotes the exactly solvable unperturbed part of the system, 
while $\{\mathcal{O}_i\}$ represent the set of terms that spoil the integrability. 
The eigenstates $\{\ket{\phi_n}\}$ of $\mathcal{H}_0$ satisfies
$\mathcal{H}_0 \ket{\phi_n} = E_n \ket{\phi_n}$, 
with the eigen energies ordered as $E_0 \leq E_1 \leq E_2 \leq \cdots$.

Within the framework of TLISA, we focus on
 \( \mathcal{H}_0 \) whose 
matrix entry $\bra{\phi_m}\mathcal{O}_i \ket{\phi_n}$
can be analytically determined. 
Then, we project the full Hamiltonian \( \mathcal{H}\) 
onto low-energy subspace spanned by states with energy cutoff $E_c = E_0 + \Lambda$,
which yields 
a truncated Hamiltonian \( \mathcal{H}_{\mathrm{trunc}} \)
with dimension of the truncated Hilbert space \( \mathcal{N} \). 
However, selecting an appropriate cutoff $\Lambda$ requires balancing computational 
cost against numerical accuracy. If $\Lambda$ is too small, truncation errors
will be significant. If $\Lambda$ is too large, the dimension $\mathcal{N}$ will exponentially 
grow for a typical lattice size $\mathcal{V} \sim 10^2$, 
making calculations too expensive. 
Moreover, \( \mathcal{H}_{\mathrm{trunc}} \) in the truncated space
spanned by $\{\ket{\phi_n}\}$ 
is typically not sparse. As a result, standard computations 
such as diagonalization or time evolution become expensive, requiring $O(\mathcal{N}^3)$ 
computation time and $O(\mathcal{N}^2)$ memory, 
and the convergence of results becomes difficult to reach.

\begin{figure}[tp]
    \includegraphics[width=8.5cm]{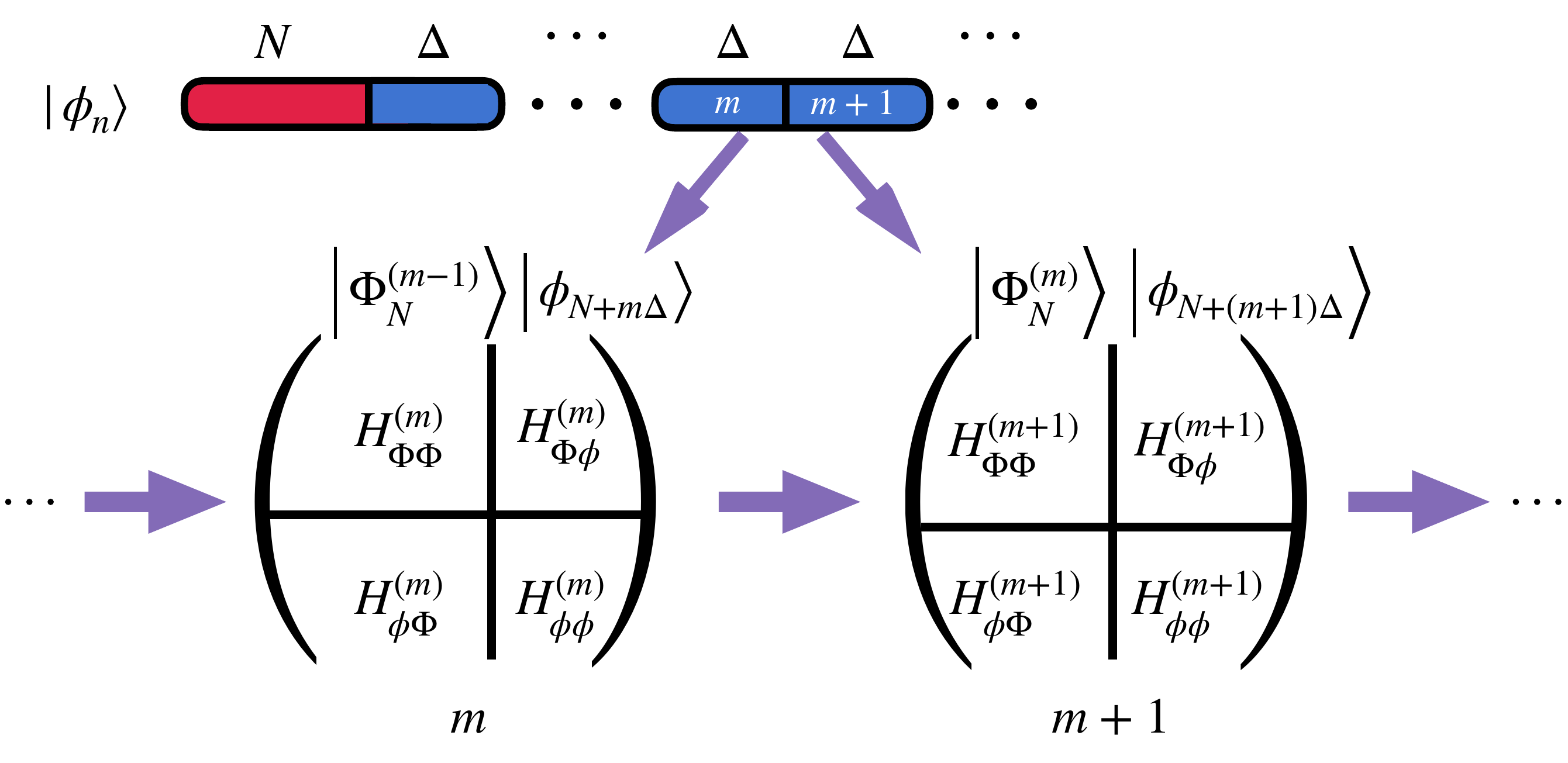}
    \caption{An illustration of the NRG procedure to diagonalize $H$ from $m$-th step to $m+1$-th step. With $H_{\Phi \Phi}^{(m)},~H_{\Phi \phi}^{(m)},~H_{\phi \Phi}^{(m)}~\text{and}~H_{\phi \phi}^{(m)}$ being the matrices block of $\mathcal{H}$ under certain bases.}
    \label{fig:NRGillustration}
\end{figure}

The NRG-TLISA is then introduced to overcome these limitations.
The machine begins by specifying an upper bound $N$ for the truncated dimension. 
In the initial step, a slightly larger basis of size $N + \Delta$ is constructed and diagonalized. 
The lowest $N$ eigenstates obtained from this diagonalization can be expressed as
\begin{equation}
\begin{aligned}
\ket{\Phi^{(1)}_{j}}&=\sum_{i=1}^{N+\Delta}\mathcal{U}^{(1)}_{ij}\ket{\phi_{i}}\\
&=\sum_{i=1}^{N}\mathcal{A}^{(1)}_{ij}\ket{\phi_{i}}+\sum_{i=1}^{\Delta}\mathcal{B}^{(1)}_{ij}\ket{\phi_{\{i+N\}}},
\label{Eq:NRG1}
\end{aligned}
\end{equation}
for $j=1,\ldots,N$,
where the unitary matrix $\mathcal{U}^{(1)}$ has been divided into $\mathcal{A}^{(1)}$ and $\mathcal{B}^{(1)}$.

The iterative procedure then proceeds as follows. From the $N + \Delta$ eigenstates obtained in the previous step, 
we keep only the lowest $N$ states $\{\ket{\Phi^{(1)}_{j}} \}$ ordered by energy, and discard the remaining $\Delta$ states. 
Then a new basis is generated from $\{\ket{\Phi^{(1)}_{j}} \}$ and the next $\Delta$ states 
$\{\ket{\phi_{N+\Delta+1}}, \cdots \ket{\phi_{N+2\Delta}} \}$ 
from $\mathcal{H}_0$, which keeps the dimension of truncated Hilbert space to be $N + \Delta$. 
Diagonalizing the Hamiltonian under the new basis yields an updated eigenstates,
\begin{equation}
\begin{aligned}
\ket{\Phi^{(2)}_{j}}&=\sum_{i=1}^{N}\mathcal{A}^{(2)}_{ij}\ket{\Phi^{(1)}_{i}}+\sum_{i=1}^{\Delta}\mathcal{B}^{(2)}_{ij}\ket{\phi_{\{i+N+\Delta\}}}\\
&=\sum_{i=1}^{N}[\mathcal{A}^{(1)}\mathcal{A}^{(2)}]_{ij}\ket{\phi_{i}}+\sum_{i=1}^{\Delta}[\mathcal{B}^{(1)}\mathcal{A}^{(2)}]_{ij}\ket{\phi_{\{i+N\}}}\\
&+\sum_{i=1}^{\Delta}\mathcal{B}^{(2)}_{ij}\ket{\phi_{\{i+N+\Delta\}}},
\label{Eq:NRG2}
\end{aligned}   
\end{equation}
for $j=1,\ldots,N$.
Repeating this iterative process for $M$ steps yields the final renormalized eigenstates:
\begin{equation}
\begin{aligned}
\ket{\Phi^{(M)}_{j}}&=\sum_{i=1}^{N}[\mathcal{A}^{(1)}\mathcal{A}^{(2)}\cdots\mathcal{A}^{(M)}]_{ij}\ket{\phi_{i}}\\
&+\sum_{i=1}^{\Delta}[\mathcal{B}^{(1)}\mathcal{A}^{(2)}\cdots\mathcal{A}^{(M)}]_{ij}\ket{\phi_{\{i+N\}}}\\
&+\cdots\\
&+\sum_{i=1}^{\Delta}\mathcal{B}^{(M)}_{ij}\ket{\phi_{\{i+N+(M-1)\Delta\}}}.
\label{Eq:NRGM}
\end{aligned}   
\end{equation}
For $j=1,\cdots,N$. The result shows that the contributions of additional $(M-1)\Delta$ states beyond the initial $N +\Delta$ 
states have been incorporated through the iterative procedures. 
The iteration continues until convergence is achieved up to a desired energy level $N$, that is,
\begin{equation}
\begin{aligned}
\forall~ l < N,~ &\big|E^{(M+1)}_{l} - E^{(M)}_{l}\big|
\leq \epsilon,
\end{aligned}
\label{Eq:NRG_convergence}
\end{equation}
where $ E^{(M)}_{l} \equiv 
\bra{\Phi^{(M)}_{l}}\mathcal{H}\ket{\Phi^{(M)}_{l}}$ and $\epsilon$ sets the accuracy goal
to control the termination of the NRG iterations [Fig.~\ref{fig:NRGillustration}]. 
In the following we demonstrate the power of the method in two 
quantum Ising models.


\textit{Examples 1.}---
As first example, we consider Ising chain with magnetic field (ICMF) of length $L$ under OBC. 
The Hamiltonian follows as 
\begin{equation}
    H_{{}_\text{ICMF}}^{{}^\text{OBC}} = -J \sum_{i=1}^{L-1}\sigma_i^{z} \sigma_{i+1}^{z}
    - J\sum_{i=1}^{L-1} \left(g \sigma_i^{x} + h_z \sigma_i^{z} \right)
    \label{eq:ising_hx_hz}
\end{equation}
with Pauli matrices $\sigma_i^\alpha$ ($\alpha = x,y,z$) at site $i$. In addition, $g$ and $h_z$ are the transverse and longitudinal field, respectively,
in units of the nearest-neighbor spin coupling $J$ between the longitudinal ($z$) components of the spins.
In the following the physical constants are fixed as $J = c = \hbar = 1$. 
The Hamiltonian
can be split into two parts $ H_{{}_\text{ICMF}}^{{}^\text{OBC}} = H_{{}_\text{TFIC}}^{{}^\text{PBC}}  + V$
where $ H_{{}_\text{TFIC}}^{{}^\text{PBC}} $ is 
the transverse field Ising chain (TFIC) with periodic boundary condition (PBC)
\begin{equation}
 H_{{}_\text{TFIC}}^{{}^\text{PBC}}  = - \sum_{i=1}^{L}\left( \sigma_i^{z} \sigma_{i+1}^{z}
    + g \sigma_i^{x} \right),~~\sigma_{L+1}^\alpha \equiv \sigma_{1}^\alpha,
\label{eq:ising_hx_PBC}
\end{equation}
and $V = \sigma_1^z \sigma_L^z-h_z \sum_{i=1}^{L} \sigma_i^{z}$.
Eq.~(\ref{eq:ising_hx_PBC}) can be 
diagonalized via the Jordan-Wigner transformation~\cite{sachdev_2011, Pfeuty}, 
while matrix entries of $V$ in eigenbasis of $ H_{{}_\text{TFIC}}^{{}^\text{PBC}} $ admit analytical expressions~\cite{Iorgov_2011}, 
which ensures the application of TLISA~\cite{rw2m-33v6}.
\begin{figure}[tp]
    \includegraphics[width = 0.998\linewidth]{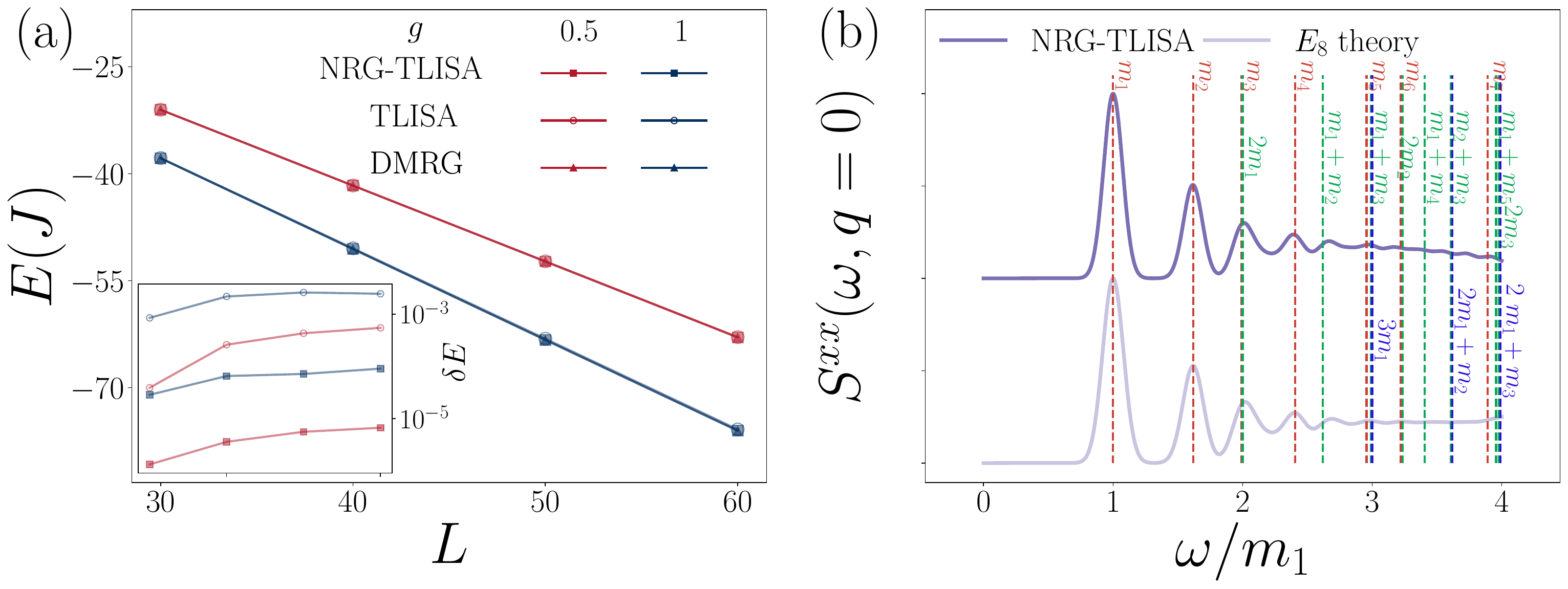}
    \caption{
    (a) Ground state energy of $H_{{}_\text{ICMF}}^{{}^\text{OBC}}$ at $h_z = 0$ for $L$ from 30 to 60, under different transverse fields with $g = 0.8$ and $g = 1.0$. The inset: $\delta E$ vs $L$ with different $g$.
    (b) DSF $S^{xx}(\omega \leq 4 m_1,q = 0)$ with $L = 100,~g = 1,~h_z = 0.05$. Red dashed lines mark masses of seven single $E_8$-particles. Green and blue dashed lines label the two-particle and three-particle threshold energies, respectively.
    }
    \label{fig:NRGChain}
\end{figure}
Fig.~\ref{fig:NRGChain}(a) compares ground state energies of $H_{{}_\text{ICMF}}^{{}^\text{OBC}}$, obtained from TLISA, NRG-TLISA, and DMRG. 
Both TLISA and NRG-TLFFA, with the truncated dimension of 512, show excellent agreement with 
high-accuracy DMRG data under maximum bond dimension $\chi_{\text{max}} = 512$. 
However, the deviation $\delta E =  \left| \frac{E_{\textrm{DMRG}} - E_{\textrm{TLISA}}}{E_{\textrm{DMRG}}} \right|$ 
demonstrated in the inset reveals a crucial distinction, where NRG-TLISA achieves two orders of magnitude better accuracy than straightforward TLISA,
indicating a substantial improvement.

Beyond energy spectrum information, our method also provides a direct access to obtain dynamical information.
Fig~\ref{fig:NRGChain}(b) displays the DSF $S^{xx}(\omega, q=0)$ for ICMF with PBC,
\begin{equation}
    S^{xx}(\omega,q = 0) = 2\pi \sum_{\mu}
    \big|\langle \psi_0 \mid \sigma_{q = 0}^x \mid \psi_\mu \rangle\big|^{2}\,
    \delta\!\big(\omega - \Delta_\mu \big).
\end{equation}
where $\sigma_{q = 0}^x = \frac{1}{\sqrt{L}} \sum_{i=1}^{L} \sigma_i^x$ and $\Delta_{\mu} = E_{\mu} - E_{\text{0}}$.
When $g=1$ and $h \to 0 $, the ICMF is described by the $E_8$ IFT in the scaling limit ($2Ja = 1$ and finite $h/a$
where $a$ is the lattice spacing constant) 
~\cite{Zamolodchikov:1989fp, PhysRevB.83.020407, Zamolodchikov:1989hfa, PhysRevB.103.235117}, 
which predicts eight massive particles with analytically known mass ratios. For comparison, Fig.~\ref{fig:NRGChain}(b) also shows 
analytical result of the DSF from the $E_8$ IFT.
Remarkably, not only the peak positions and their relative amplitudes, but also the overall line shape, show excellent agreement with analytical results, indicating that our framework accurately captures the system's low-energy physics.

\textit{Examples 2.}---
The second example considers two coupled TFICs with inter-chain Ising coupling, described by
\begin{equation}
    H_{\text{II}} = H^{(1)}_{{}_\text{TFIC}} + H^{(2)}_{{}_\text{TFIC}} - g_i \sum_{j = 1}^{L} \sigma^{z(1)}_j \sigma^{z(2)}_j,
\end{equation}
where $H^{(\alpha)}_{{}_\text{TFIC}},~\alpha = 1,2$ denotes the Hamiltonian of a ferromagnetic TFIC with OBC,
\begin{equation}
    H^{(\alpha)}_{{}_{\text{TFIC}}} = -\sum_{i=1}^{L-1} \sigma_i^{z(\alpha)}\sigma_{i+1}^{z(\alpha)} - g\sum_{i=1}^{L} \sigma_i^{x(\alpha)}.
    \label{Eq:Ising_ladder}
\end{equation}
Each chain of the ladder is first solved within the TLISA framework based on its integrable part, yielding eigenstates of $H^{(\alpha)}_{0}$~($\alpha=1,2$) as the initial basis. Applying NRG to each chain produces low-energy renormalized states $\{\ket{\Phi_n}\}_{n=1}^{N}$. 
The ladder Hilbert space is then spanned by tensor products $\{\ket{\Phi_m}\otimes\ket{\Phi_n}\}_{m,n=1}^{N}$, ordered by energy to build the low-energy subspace, on which the NRG iteration is performed again to efficiently capture the ladder's low-energy physics.

\begin{figure}[tp]
    \includegraphics[width=8.5cm]{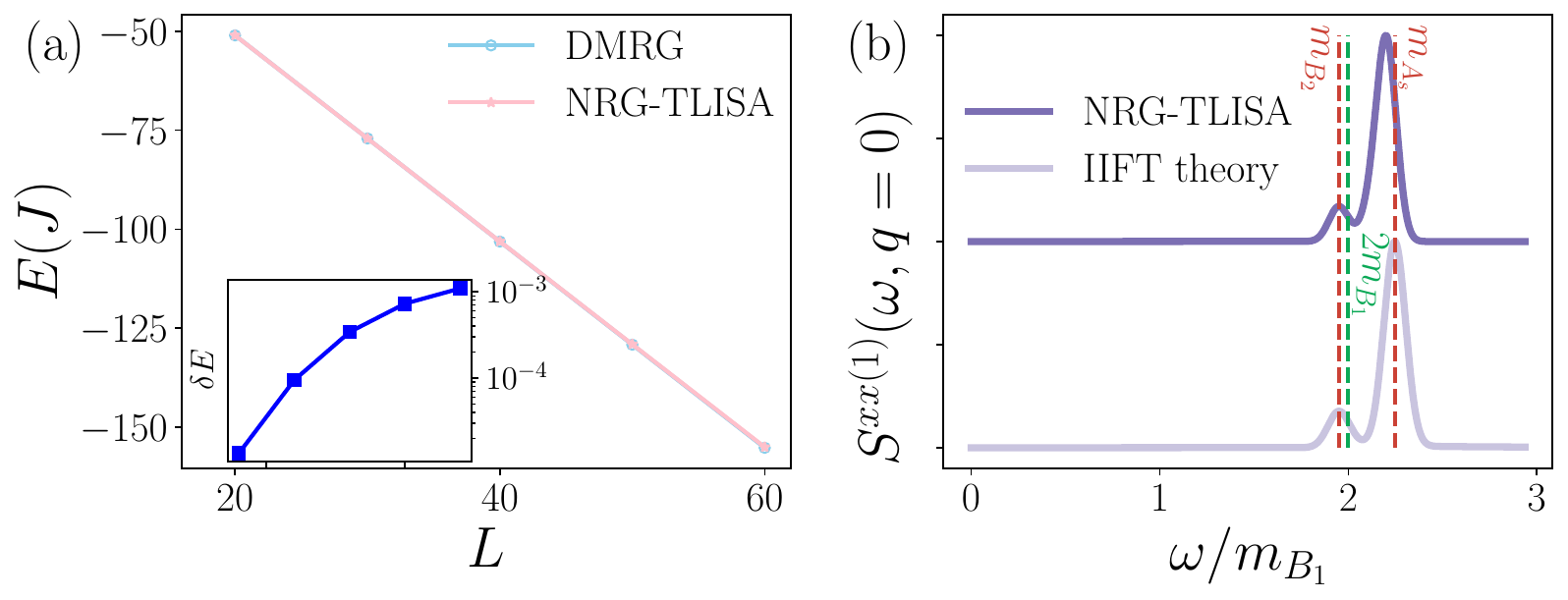}
    \caption{
        (a) Ground state energy of $H_{\text{II}}$ with OBC calculated by DMRG and NRG-TLISA for size $L = 20$ to $L=60$ with $g = 1,~~g_i = 0.1$.
        Inset: $\delta E$ vs $L$.
        (b) DSF $S^{xx(1)}(\omega<3m_{B_1},q = 0)$ with $L = 100,~g = 1,~g_i = 0.1$. Red dashed lines denote the breather $m_{B_2}$ and soliton $m_A$ masses and green dashed lines denote the two-particle~($2 m_{B_1}$) threshold energy.
    }
    \label{fig:LadderDsf}
\end{figure}
 
The Hilbert space of each chain is truncated to 2048 renormalized states. During NRG on the ladder subspace, we then retain 4096 states per iteration, yielding convergence of the lowest 256 eigenstates.
DMRG calculations employ a maximum bond dimension $\chi_{\text{max}} = 512$[Fig.~\ref{fig:LadderDsf}(a)].
Excellent agreement between the two methods confirms the 
accuracy of the TLISA-NRG approach for ladder systems.
Fig.~\ref{fig:LadderDsf}(b) presents the DSF $S^{xx(1)}(\omega,q = 0)$ under PBC where the subscript $(1)$ indicates that the operator $\sigma^x$ 
of chain $(1)$ only.

At $g = 1$, the continuum (scaling) limit of Eq.~(\ref{Eq:Ising_ladder}) is described by the $\text{Ising}_{h}^{2}$ integrable field theory (IIFT)~\cite{LeClair:1997gv,Gao:2024ngb,Gao:2025mcg}
\begin{equation}
    \mathcal{H}_{\text{II}} = \mathcal{H}^{(1)}_{c = 1/2} + \mathcal{H}^{(2)}_{c = 1/2} + \lambda \int dx \sigma^{(1)}(x) \sigma^{(2)}(x),
\end{equation}
Here, $\mathcal{H}^{(i)}_{c = 1/2}$ $(i=1,\;2)$ represents the CFT Hamiltonian with central charge $c = 1/2$ for chain $(i)$, 
and $\lambda$ denotes the rescaled interchain coupling. The corresponding $S^{xx(1)}(\omega, q = 0)$ from the IIFT is plotted in Fig.~\ref{fig:LadderDsf}(b), showing excellent agreement with our numerical results. 
Furthermore, as shown in Fig.~\ref{fig:LadderDsf}(b), no distinct peak corresponding to the breather $m_{B_1}$ is observed, which is consistent with the
"dark particle" predictions of the IIFT~\cite{Gao:2024ngb}. 
This overall agreement demonstrates that the proposed method efficiently captures the low-energy physics of ladder systems.

\textit{Discussions and conclusions.}--- 
Both TLISA and NRG-TLISA rest on an assumption that when the eigenstates 
of the integrable part \( \mathcal{H}_0 \) are ordered by energy, the low-energy behavior
of the full Hamiltonian \( \mathcal{H} \) is governed by the lowest-lying eigenstates of \( \mathcal{H}_0 \). 
From the renormalization group (RG) perspective, the assumption holds for relevant terms whose couplings increase under the RG flow, 
thereby enhancing low-energy mixing, suppressing high-energy contributions, 
and hence allowing a modest cutoff \(E_{\mathrm{cut}}\) to capture the essential physics. 
The scenario changes if the additional couplings are marginal or irrelevant. 
Their leading 
influence lies at higher energies, so comparable accuracy requires a much larger \( E_{\mathrm{cut}} \). 
The retained Hilbert space then increases exponentially, and the method loses efficiency.

Overall, the NRG-TLISA hybrid method extends the NRG-TSA framework of Konik and Adamov~\cite{PhysRevLett.98.147205, Brandino:2010sv}. The original framework targets continuum models based on CFT and IFT, whereas our approach applies to lattice models.
It serves as a practical computational tool and also as a methodological advance for 
strongly correlated quantum systems. By combining the systematic procedures of NRG with the truncation strategy of TLISA, 
the framework reaches a trade-off between accuracy and scalability.
For the ICMF model, our framework agrees closely with DMRG at considerable system sizes. This indicates that it can handle systems beyond the size range accessible to ED.
In addition, the method provides the information of wavefunctions for multiple eigenstates. 
In contrast, DMRG typically approximately gives only ground-state and a few low-lying excited states. Therefore, we can directly calculate DSF, which is consistent with analytical predictions. 
We further study ladder models and find ground state energies in excellent agreement with DMRG, 
and the DSF also agree well with analytical results. 

As demonstrated in this work, our framework can straightforwardly apply to 
non-integrable systems built on
integrable ones, namely, generic deformation of integrable 
lattice models. It
not only yields the eigenenergies and DSF of 
systems but also provides accurately renormalized truncated wavefunctions. 
This 
enables analyses of entanglement entropy for low-lying excited states and study on out-of-equilibrium dynamics. 
It would be worthwhile to further apply our method to systems built on Bethe-ansatz-solvable 
models (e.g. the XXZ~\cite{Takhtajan:1979iv} and Lieb-Liniger~\cite{PhysRevLett.19.1312} models) for which analytical form factors are available~\cite{KITANINE1999647, slavnov_1990}. 
Furthermore, our approach offers a possible route for exploring two-dimensional quantum many-body systems
built on integrable models.

\textit{Acknowledgements.}--- We thank R. Yu, D. Yang, X. Yu, Y. Jiang, Y. Gao, N. Xi, E. Lv, and Q. Tang 
for helpful discussions.
This work is supported by the National Natural Science Foundation of China Nos. 12274288, 12450004, 
the Innovation Program for Quantum Science and Technology Grant No. 2021ZD0301900.

\bibliography{main}

\begin{thebibliography}{65}%
\makeatletter
\providecommand \@ifxundefined [1]{%
 \@ifx{#1\undefined}
}%
\providecommand \@ifnum [1]{%
 \ifnum #1\expandafter \@firstoftwo
 \else \expandafter \@secondoftwo
 \fi
}%
\providecommand \@ifx [1]{%
 \ifx #1\expandafter \@firstoftwo
 \else \expandafter \@secondoftwo
 \fi
}%
\providecommand \natexlab [1]{#1}%
\providecommand \enquote  [1]{``#1''}%
\providecommand \bibnamefont  [1]{#1}%
\providecommand \bibfnamefont [1]{#1}%
\providecommand \citenamefont [1]{#1}%
\providecommand \href@noop [0]{\@secondoftwo}%
\providecommand \href [0]{\begingroup \@sanitize@url \@href}%
\providecommand \@href[1]{\@@startlink{#1}\@@href}%
\providecommand \@@href[1]{\endgroup#1\@@endlink}%
\providecommand \@sanitize@url [0]{\catcode `\\12\catcode `\$12\catcode
  `\&12\catcode `\#12\catcode `\^12\catcode `\_12\catcode `\%12\relax}%
\providecommand \@@startlink[1]{}%
\providecommand \@@endlink[0]{}%
\providecommand \url  [0]{\begingroup\@sanitize@url \@url }%
\providecommand \@url [1]{\endgroup\@href {#1}{\urlprefix }}%
\providecommand \urlprefix  [0]{URL }%
\providecommand \Eprint [0]{\href }%
\providecommand \doibase [0]{https://doi.org/}%
\providecommand \selectlanguage [0]{\@gobble}%
\providecommand \bibinfo  [0]{\@secondoftwo}%
\providecommand \bibfield  [0]{\@secondoftwo}%
\providecommand \translation [1]{[#1]}%
\providecommand \BibitemOpen [0]{}%
\providecommand \bibitemStop [0]{}%
\providecommand \bibitemNoStop [0]{.\EOS\space}%
\providecommand \EOS [0]{\spacefactor3000\relax}%
\providecommand \BibitemShut  [1]{\csname bibitem#1\endcsname}%
\let\auto@bib@innerbib\@empty
\bibitem [{\citenamefont {Bethe}(1931)}]{Bethe}%
  \BibitemOpen
  \bibfield  {author} {\bibinfo {author} {\bibfnamefont {H.}~\bibnamefont
  {Bethe}},\ }\bibfield  {title} {\bibinfo {title} {Zur theorie der metalle},\
  }\href {https://doi.org/10.1007/BF01341708} {\bibfield  {journal} {\bibinfo
  {journal} {Z. Phys.}\ }\textbf {\bibinfo {volume} {71}},\ \bibinfo {pages}
  {205} (\bibinfo {year} {1931})}\BibitemShut {NoStop}%
\bibitem [{\citenamefont {Guan}\ \emph {et~al.}(2013)\citenamefont {Guan},
  \citenamefont {Batchelor},\ and\ \citenamefont {Lee}}]{RevModPhys.85.1633}%
  \BibitemOpen
  \bibfield  {author} {\bibinfo {author} {\bibfnamefont {X.-W.}\ \bibnamefont
  {Guan}}, \bibinfo {author} {\bibfnamefont {M.~T.}\ \bibnamefont
  {Batchelor}},\ and\ \bibinfo {author} {\bibfnamefont {C.}~\bibnamefont
  {Lee}},\ }\bibfield  {title} {\bibinfo {title} {Fermi gases in one dimension:
  From {B}ethe ansatz to experiments},\ }\href
  {https://doi.org/10.1103/RevModPhys.85.1633} {\bibfield  {journal} {\bibinfo
  {journal} {Rev. Mod. Phys.}\ }\textbf {\bibinfo {volume} {85}},\ \bibinfo
  {pages} {1633} (\bibinfo {year} {2013})}\BibitemShut {NoStop}%
\bibitem [{\citenamefont {He}\ \emph {et~al.}(2017)\citenamefont {He},
  \citenamefont {Jiang}, \citenamefont {Yu}, \citenamefont {Lin},\ and\
  \citenamefont {Guan}}]{PhysRevB.96.220401}%
  \BibitemOpen
  \bibfield  {author} {\bibinfo {author} {\bibfnamefont {F.}~\bibnamefont
  {He}}, \bibinfo {author} {\bibfnamefont {Y.}~\bibnamefont {Jiang}}, \bibinfo
  {author} {\bibfnamefont {Y.-C.}\ \bibnamefont {Yu}}, \bibinfo {author}
  {\bibfnamefont {H.-Q.}\ \bibnamefont {Lin}},\ and\ \bibinfo {author}
  {\bibfnamefont {X.-W.}\ \bibnamefont {Guan}},\ }\bibfield  {title} {\bibinfo
  {title} {Quantum criticality of spinons},\ }\href
  {https://doi.org/10.1103/PhysRevB.96.220401} {\bibfield  {journal} {\bibinfo
  {journal} {Phys. Rev. B}\ }\textbf {\bibinfo {volume} {96}},\ \bibinfo
  {pages} {220401} (\bibinfo {year} {2017})}\BibitemShut {NoStop}%
\bibitem [{\citenamefont {Takahashi}\ and\ \citenamefont
  {Suzuki}(1972)}]{10.1143PTP.48.2187}%
  \BibitemOpen
  \bibfield  {author} {\bibinfo {author} {\bibfnamefont {M.}~\bibnamefont
  {Takahashi}}\ and\ \bibinfo {author} {\bibfnamefont {M.}~\bibnamefont
  {Suzuki}},\ }\bibfield  {title} {\bibinfo {title} {One-dimensional
  anisotropic {H}eisenberg model at finite temperatures},\ }\href
  {https://doi.org/10.1143/PTP.48.2187} {\bibfield  {journal} {\bibinfo
  {journal} {Prog. Theor. Phys.}\ }\textbf {\bibinfo {volume} {48}},\ \bibinfo
  {pages} {2187} (\bibinfo {year} {1972})}\BibitemShut {NoStop}%
\bibitem [{\citenamefont {Haldane}(1983)}]{PhysRevLett.51.605}%
  \BibitemOpen
  \bibfield  {author} {\bibinfo {author} {\bibfnamefont {F.~D.~M.}\
  \bibnamefont {Haldane}},\ }\bibfield  {title} {\bibinfo {title} {Fractional
  quantization of the {H}all effect: A hierarchy of incompressible quantum
  fluid states},\ }\href {https://doi.org/10.1103/PhysRevLett.51.605}
  {\bibfield  {journal} {\bibinfo  {journal} {Phys. Rev. Lett.}\ }\textbf
  {\bibinfo {volume} {51}},\ \bibinfo {pages} {605} (\bibinfo {year}
  {1983})}\BibitemShut {NoStop}%
\bibitem [{\citenamefont {Laughlin}(1983)}]{PhysRevLett.50.1395}%
  \BibitemOpen
  \bibfield  {author} {\bibinfo {author} {\bibfnamefont {R.~B.}\ \bibnamefont
  {Laughlin}},\ }\bibfield  {title} {\bibinfo {title} {{Anomalous quantum Hall
  effect: An incompressible quantum fluid with fractionally charged
  excitations}},\ }\href {https://doi.org/10.1103/PhysRevLett.50.1395}
  {\bibfield  {journal} {\bibinfo  {journal} {Phys. Rev. Lett.}\ }\textbf
  {\bibinfo {volume} {50}},\ \bibinfo {pages} {1395} (\bibinfo {year}
  {1983})}\BibitemShut {NoStop}%
\bibitem [{\citenamefont {Bulla}\ \emph {et~al.}(2008)\citenamefont {Bulla},
  \citenamefont {Costi},\ and\ \citenamefont {Pruschke}}]{RevModPhys.80.395}%
  \BibitemOpen
  \bibfield  {author} {\bibinfo {author} {\bibfnamefont {R.}~\bibnamefont
  {Bulla}}, \bibinfo {author} {\bibfnamefont {T.~A.}\ \bibnamefont {Costi}},\
  and\ \bibinfo {author} {\bibfnamefont {T.}~\bibnamefont {Pruschke}},\
  }\bibfield  {title} {\bibinfo {title} {Numerical renormalization group method
  for quantum impurity systems},\ }\href
  {https://doi.org/10.1103/RevModPhys.80.395} {\bibfield  {journal} {\bibinfo
  {journal} {Rev. Mod. Phys.}\ }\textbf {\bibinfo {volume} {80}},\ \bibinfo
  {pages} {395} (\bibinfo {year} {2008})}\BibitemShut {NoStop}%
\bibitem [{\citenamefont {Bulla}\ \emph {et~al.}(2001)\citenamefont {Bulla},
  \citenamefont {Costi},\ and\ \citenamefont {Vollhardt}}]{PhysRevB.64.045103}%
  \BibitemOpen
  \bibfield  {author} {\bibinfo {author} {\bibfnamefont {R.}~\bibnamefont
  {Bulla}}, \bibinfo {author} {\bibfnamefont {T.~A.}\ \bibnamefont {Costi}},\
  and\ \bibinfo {author} {\bibfnamefont {D.}~\bibnamefont {Vollhardt}},\
  }\bibfield  {title} {\bibinfo {title} {Finite-temperature numerical
  renormalization group study of the {M}ott transition},\ }\href
  {https://doi.org/10.1103/PhysRevB.64.045103} {\bibfield  {journal} {\bibinfo
  {journal} {Phys. Rev. B}\ }\textbf {\bibinfo {volume} {64}},\ \bibinfo
  {pages} {045103} (\bibinfo {year} {2001})}\BibitemShut {NoStop}%
\bibitem [{\citenamefont {Peters}\ \emph {et~al.}(2006)\citenamefont {Peters},
  \citenamefont {Pruschke},\ and\ \citenamefont {Anders}}]{PhysRevB.74.245114}%
  \BibitemOpen
  \bibfield  {author} {\bibinfo {author} {\bibfnamefont {R.}~\bibnamefont
  {Peters}}, \bibinfo {author} {\bibfnamefont {T.}~\bibnamefont {Pruschke}},\
  and\ \bibinfo {author} {\bibfnamefont {F.~B.}\ \bibnamefont {Anders}},\
  }\bibfield  {title} {\bibinfo {title} {Numerical renormalization group
  approach to {G}reen's functions for quantum impurity models},\ }\href
  {https://doi.org/10.1103/PhysRevB.74.245114} {\bibfield  {journal} {\bibinfo
  {journal} {Phys. Rev. B}\ }\textbf {\bibinfo {volume} {74}},\ \bibinfo
  {pages} {245114} (\bibinfo {year} {2006})}\BibitemShut {NoStop}%
\bibitem [{\citenamefont {Weichselbaum}\ and\ \citenamefont {von
  Delft}(2007)}]{PhysRevLett.99.076402}%
  \BibitemOpen
  \bibfield  {author} {\bibinfo {author} {\bibfnamefont {A.}~\bibnamefont
  {Weichselbaum}}\ and\ \bibinfo {author} {\bibfnamefont {J.}~\bibnamefont {von
  Delft}},\ }\bibfield  {title} {\bibinfo {title} {Sum-rule conserving spectral
  functions from the numerical renormalization group},\ }\href
  {https://doi.org/10.1103/PhysRevLett.99.076402} {\bibfield  {journal}
  {\bibinfo  {journal} {Phys. Rev. Lett.}\ }\textbf {\bibinfo {volume} {99}},\
  \bibinfo {pages} {076402} (\bibinfo {year} {2007})}\BibitemShut {NoStop}%
\bibitem [{\citenamefont {White}(1992)}]{PhysRevLett.69.2863}%
  \BibitemOpen
  \bibfield  {author} {\bibinfo {author} {\bibfnamefont {S.~R.}\ \bibnamefont
  {White}},\ }\bibfield  {title} {\bibinfo {title} {Density matrix formulation
  for quantum renormalization groups},\ }\href
  {https://doi.org/10.1103/PhysRevLett.69.2863} {\bibfield  {journal} {\bibinfo
   {journal} {Phys. Rev. Lett.}\ }\textbf {\bibinfo {volume} {69}},\ \bibinfo
  {pages} {2863} (\bibinfo {year} {1992})}\BibitemShut {NoStop}%
\bibitem [{\citenamefont {Schollw\"ock}(2011)}]{SCHOLLWOCK201196}%
  \BibitemOpen
  \bibfield  {author} {\bibinfo {author} {\bibfnamefont {U.}~\bibnamefont
  {Schollw\"ock}},\ }\bibfield  {title} {\bibinfo {title} {The density-matrix
  renormalization group in the age of matrix product states},\ }\href
  {https://doi.org/https://doi.org/10.1016/j.aop.2010.09.012} {\bibfield
  {journal} {\bibinfo  {journal} {Ann. Phys.}\ }\textbf {\bibinfo {volume}
  {326}},\ \bibinfo {pages} {96} (\bibinfo {year} {2011})}\BibitemShut
  {NoStop}%
\bibitem [{\citenamefont {Cirac}\ \emph {et~al.}(2021)\citenamefont {Cirac},
  \citenamefont {P\'erez-Garc\'{\i}a}, \citenamefont {Schuch},\ and\
  \citenamefont {Verstraete}}]{RevModPhys.93.045003}%
  \BibitemOpen
  \bibfield  {author} {\bibinfo {author} {\bibfnamefont {J.~I.}\ \bibnamefont
  {Cirac}}, \bibinfo {author} {\bibfnamefont {D.}~\bibnamefont
  {P\'erez-Garc\'{\i}a}}, \bibinfo {author} {\bibfnamefont {N.}~\bibnamefont
  {Schuch}},\ and\ \bibinfo {author} {\bibfnamefont {F.}~\bibnamefont
  {Verstraete}},\ }\bibfield  {title} {\bibinfo {title} {Matrix product states
  and projected entangled pair states: Concepts, symmetries, theorems},\ }\href
  {https://doi.org/10.1103/RevModPhys.93.045003} {\bibfield  {journal}
  {\bibinfo  {journal} {Rev. Mod. Phys.}\ }\textbf {\bibinfo {volume} {93}},\
  \bibinfo {pages} {045003} (\bibinfo {year} {2021})}\BibitemShut {NoStop}%
\bibitem [{\citenamefont {Or\'us}(2014)}]{Orus:2013kga}%
  \BibitemOpen
  \bibfield  {author} {\bibinfo {author} {\bibfnamefont {R.}~\bibnamefont
  {Or\'us}},\ }\bibfield  {title} {\bibinfo {title} {A practical introduction
  to tensor networks: Matrix product states and projected entangled pair
  states},\ }\href {https://doi.org/https://doi.org/10.1016/j.aop.2014.06.013}
  {\bibfield  {journal} {\bibinfo  {journal} {Ann. Phys.}\ }\textbf {\bibinfo
  {volume} {349}},\ \bibinfo {pages} {117} (\bibinfo {year}
  {2014})}\BibitemShut {NoStop}%
\bibitem [{\citenamefont {Foulkes}\ \emph {et~al.}(2001)\citenamefont
  {Foulkes}, \citenamefont {Mitas}, \citenamefont {Needs},\ and\ \citenamefont
  {Rajagopal}}]{RevModPhys.73.33}%
  \BibitemOpen
  \bibfield  {author} {\bibinfo {author} {\bibfnamefont {W.~M.~C.}\
  \bibnamefont {Foulkes}}, \bibinfo {author} {\bibfnamefont {L.}~\bibnamefont
  {Mitas}}, \bibinfo {author} {\bibfnamefont {R.~J.}\ \bibnamefont {Needs}},\
  and\ \bibinfo {author} {\bibfnamefont {G.}~\bibnamefont {Rajagopal}},\
  }\bibfield  {title} {\bibinfo {title} {Quantum {M}onte {C}arlo simulations of
  solids},\ }\href {https://doi.org/10.1103/RevModPhys.73.33} {\bibfield
  {journal} {\bibinfo  {journal} {Rev. Mod. Phys.}\ }\textbf {\bibinfo {volume}
  {73}},\ \bibinfo {pages} {33} (\bibinfo {year} {2001})}\BibitemShut {NoStop}%
\bibitem [{\citenamefont {Blankenbecler}\ \emph {et~al.}(1981)\citenamefont
  {Blankenbecler}, \citenamefont {Scalapino},\ and\ \citenamefont
  {Sugar}}]{PhysRevD.24.2278}%
  \BibitemOpen
  \bibfield  {author} {\bibinfo {author} {\bibfnamefont {R.}~\bibnamefont
  {Blankenbecler}}, \bibinfo {author} {\bibfnamefont {D.~J.}\ \bibnamefont
  {Scalapino}},\ and\ \bibinfo {author} {\bibfnamefont {R.~L.}\ \bibnamefont
  {Sugar}},\ }\bibfield  {title} {\bibinfo {title} {Monte {C}arlo calculations
  of coupled boson-fermion systems. {I}},\ }\href
  {https://doi.org/10.1103/PhysRevD.24.2278} {\bibfield  {journal} {\bibinfo
  {journal} {Phys. Rev. D}\ }\textbf {\bibinfo {volume} {24}},\ \bibinfo
  {pages} {2278} (\bibinfo {year} {1981})}\BibitemShut {NoStop}%
\bibitem [{\citenamefont {Troyer}\ and\ \citenamefont
  {Wiese}(2005)}]{PhysRevLett.94.170201}%
  \BibitemOpen
  \bibfield  {author} {\bibinfo {author} {\bibfnamefont {M.}~\bibnamefont
  {Troyer}}\ and\ \bibinfo {author} {\bibfnamefont {U.-J.}\ \bibnamefont
  {Wiese}},\ }\bibfield  {title} {\bibinfo {title} {Computational complexity
  and fundamental limitations to fermionic quantum {M}onte {C}arlo
  simulations},\ }\href {https://doi.org/10.1103/PhysRevLett.94.170201}
  {\bibfield  {journal} {\bibinfo  {journal} {Phys. Rev. Lett.}\ }\textbf
  {\bibinfo {volume} {94}},\ \bibinfo {pages} {170201} (\bibinfo {year}
  {2005})}\BibitemShut {NoStop}%
\bibitem [{\citenamefont {Alexandru}\ \emph {et~al.}(2022)\citenamefont
  {Alexandru}, \citenamefont {Ba{\c s}ar}, \citenamefont {Bedaque},\ and\
  \citenamefont {Warrington}}]{RevModPhys.94.015006}%
  \BibitemOpen
  \bibfield  {author} {\bibinfo {author} {\bibfnamefont {A.}~\bibnamefont
  {Alexandru}}, \bibinfo {author} {\bibfnamefont {G.}~\bibnamefont {Ba{\c
  s}ar}}, \bibinfo {author} {\bibfnamefont {P.~F.}\ \bibnamefont {Bedaque}},\
  and\ \bibinfo {author} {\bibfnamefont {N.~C.}\ \bibnamefont {Warrington}},\
  }\bibfield  {title} {\bibinfo {title} {Complex paths around the sign
  problem},\ }\href {https://doi.org/10.1103/RevModPhys.94.015006} {\bibfield
  {journal} {\bibinfo  {journal} {Rev. Mod. Phys.}\ }\textbf {\bibinfo {volume}
  {94}},\ \bibinfo {pages} {015006} (\bibinfo {year} {2022})}\BibitemShut
  {NoStop}%
\bibitem [{\citenamefont {Zhang}\ and\ \citenamefont
  {Kim}(2017)}]{PhysRevLett.118.216401}%
  \BibitemOpen
  \bibfield  {author} {\bibinfo {author} {\bibfnamefont {Y.}~\bibnamefont
  {Zhang}}\ and\ \bibinfo {author} {\bibfnamefont {E.-A.}\ \bibnamefont
  {Kim}},\ }\bibfield  {title} {\bibinfo {title} {Quantum loop topography for
  machine learning},\ }\href {https://doi.org/10.1103/PhysRevLett.118.216401}
  {\bibfield  {journal} {\bibinfo  {journal} {Phys. Rev. Lett.}\ }\textbf
  {\bibinfo {volume} {118}},\ \bibinfo {pages} {216401} (\bibinfo {year}
  {2017})}\BibitemShut {NoStop}%
\bibitem [{\citenamefont {Chen}\ \emph {et~al.}(2018)\citenamefont {Chen},
  \citenamefont {Cheng}, \citenamefont {Xie}, \citenamefont {Wang},\ and\
  \citenamefont {Xiang}}]{PhysRevB.97.085104}%
  \BibitemOpen
  \bibfield  {author} {\bibinfo {author} {\bibfnamefont {J.}~\bibnamefont
  {Chen}}, \bibinfo {author} {\bibfnamefont {S.}~\bibnamefont {Cheng}},
  \bibinfo {author} {\bibfnamefont {H.}~\bibnamefont {Xie}}, \bibinfo {author}
  {\bibfnamefont {L.}~\bibnamefont {Wang}},\ and\ \bibinfo {author}
  {\bibfnamefont {T.}~\bibnamefont {Xiang}},\ }\bibfield  {title} {\bibinfo
  {title} {Equivalence of restricted {B}oltzmann machines and tensor network
  states},\ }\href {https://doi.org/10.1103/PhysRevB.97.085104} {\bibfield
  {journal} {\bibinfo  {journal} {Phys. Rev. B}\ }\textbf {\bibinfo {volume}
  {97}},\ \bibinfo {pages} {085104} (\bibinfo {year} {2018})}\BibitemShut
  {NoStop}%
\bibitem [{\citenamefont {Carrasquilla}\ and\ \citenamefont
  {Torlai}(2021)}]{PRXQuantum.2.040201}%
  \BibitemOpen
  \bibfield  {author} {\bibinfo {author} {\bibfnamefont {J.}~\bibnamefont
  {Carrasquilla}}\ and\ \bibinfo {author} {\bibfnamefont {G.}~\bibnamefont
  {Torlai}},\ }\bibfield  {title} {\bibinfo {title} {How to use neural networks
  to investigate quantum many-body physics},\ }\href
  {https://doi.org/10.1103/PRXQuantum.2.040201} {\bibfield  {journal} {\bibinfo
   {journal} {PRX Quantum}\ }\textbf {\bibinfo {volume} {2}},\ \bibinfo {pages}
  {040201} (\bibinfo {year} {2021})}\BibitemShut {NoStop}%
\bibitem [{\citenamefont {Yurov}\ and\ \citenamefont
  {Zamolodchikov}(1990)}]{Yurov:1989yu}%
  \BibitemOpen
  \bibfield  {author} {\bibinfo {author} {\bibfnamefont {V.~P.}\ \bibnamefont
  {Yurov}}\ and\ \bibinfo {author} {\bibfnamefont {A.~B.}\ \bibnamefont
  {Zamolodchikov}},\ }\bibfield  {title} {\bibinfo {title} {{Truncated
  conformal space approach to scaling Lee-Yang model}},\ }\href
  {https://doi.org/10.1142/S0217751X9000218X} {\bibfield  {journal} {\bibinfo
  {journal} {Int. J. Mod. Phys. A}\ }\textbf {\bibinfo {volume} {5}},\ \bibinfo
  {pages} {3221} (\bibinfo {year} {1990})}\BibitemShut {NoStop}%
\bibitem [{\citenamefont {Yurov}\ and\ \citenamefont
  {Zamolodchikov}(1991{\natexlab{a}})}]{Yurov:1991my}%
  \BibitemOpen
  \bibfield  {author} {\bibinfo {author} {\bibfnamefont {V.~P.}\ \bibnamefont
  {Yurov}}\ and\ \bibinfo {author} {\bibfnamefont {A.~B.}\ \bibnamefont
  {Zamolodchikov}},\ }\bibfield  {title} {\bibinfo {title} {{Truncated
  fermionic space approach to the critical 2D Ising model with magnetic
  field}},\ }\href {https://doi.org/10.1142/S0217751X91002161} {\bibfield
  {journal} {\bibinfo  {journal} {Int. J. Mod. Phys. A}\ }\textbf {\bibinfo
  {volume} {6}},\ \bibinfo {pages} {4557} (\bibinfo {year}
  {1991}{\natexlab{a}})}\BibitemShut {NoStop}%
\bibitem [{\citenamefont {Fitzpatrick}\ and\ \citenamefont
  {Katz}(2022)}]{fitzpatrick2022snowmasswhitepaperhamiltonian}%
  \BibitemOpen
  \bibfield  {author} {\bibinfo {author} {\bibfnamefont {A.~L.}\ \bibnamefont
  {Fitzpatrick}}\ and\ \bibinfo {author} {\bibfnamefont {E.}~\bibnamefont
  {Katz}},\ }\bibfield  {title} {\bibinfo {title} {Snowmass white paper:
  Hamiltonian truncation},\ }\href {https://arxiv.org/abs/2201.11696}
  {\bibfield  {journal} {\bibinfo  {journal} {arXiv: 2201.11696}\ } (\bibinfo
  {year} {2022})}\BibitemShut {NoStop}%
\bibitem [{\citenamefont {Hogervorst}\ \emph {et~al.}(2015)\citenamefont
  {Hogervorst}, \citenamefont {Rychkov},\ and\ \citenamefont {van
  Rees}}]{PhysRevD.91.025005}%
  \BibitemOpen
  \bibfield  {author} {\bibinfo {author} {\bibfnamefont {M.}~\bibnamefont
  {Hogervorst}}, \bibinfo {author} {\bibfnamefont {S.}~\bibnamefont
  {Rychkov}},\ and\ \bibinfo {author} {\bibfnamefont {B.~C.}\ \bibnamefont {van
  Rees}},\ }\bibfield  {title} {\bibinfo {title} {Truncated conformal space
  approach in $d$ dimensions: A cheap alternative to lattice field theory?},\
  }\href {https://doi.org/10.1103/PhysRevD.91.025005} {\bibfield  {journal}
  {\bibinfo  {journal} {Phys. Rev. D}\ }\textbf {\bibinfo {volume} {91}},\
  \bibinfo {pages} {025005} (\bibinfo {year} {2015})}\BibitemShut {NoStop}%
\bibitem [{\citenamefont {Mir{\'o}}\ and\ \citenamefont
  {Ingoldby}(2022)}]{EliasMiro:2021aof}%
  \BibitemOpen
  \bibfield  {author} {\bibinfo {author} {\bibfnamefont {J.~E.}\ \bibnamefont
  {Mir{\'o}}}\ and\ \bibinfo {author} {\bibfnamefont {J.}~\bibnamefont
  {Ingoldby}},\ }\bibfield  {title} {\bibinfo {title} {Hamiltonian truncation
  with larger dimensions},\ }\href {https://doi.org/10.1007/JHEP05(2022)151}
  {\bibfield  {journal} {\bibinfo  {journal} {J. High Energy Phys.}\ }\textbf
  {\bibinfo {volume} {2022}}\bibinfo  {number} { (5)},\ \bibinfo {pages}
  {151}}\BibitemShut {NoStop}%
\bibitem [{\citenamefont {Yurov}\ and\ \citenamefont
  {Zamolodchikov}(1991{\natexlab{b}})}]{Yurov:1990kv}%
  \BibitemOpen
\bibfield  {number} {  }\bibfield  {author} {\bibinfo {author} {\bibfnamefont
  {V.~P.}\ \bibnamefont {Yurov}}\ and\ \bibinfo {author} {\bibfnamefont
  {A.~B.}\ \bibnamefont {Zamolodchikov}},\ }\bibfield  {title} {\bibinfo
  {title} {{Correlation functions of integrable 2D models of relativistic field
  theory. Ising model}},\ }\href {https://doi.org/10.1142/S0217751X91001660}
  {\bibfield  {journal} {\bibinfo  {journal} {Int. J. Mod. Phys. A}\ }\textbf
  {\bibinfo {volume} {6}},\ \bibinfo {pages} {3419} (\bibinfo {year}
  {1991}{\natexlab{b}})}\BibitemShut {NoStop}%
\bibitem [{\citenamefont {Lassig}\ \emph {et~al.}(1991)\citenamefont {Lassig},
  \citenamefont {Mussardo},\ and\ \citenamefont {Cardy}}]{Lassig:1990xy}%
  \BibitemOpen
  \bibfield  {author} {\bibinfo {author} {\bibfnamefont {M.}~\bibnamefont
  {Lassig}}, \bibinfo {author} {\bibfnamefont {G.}~\bibnamefont {Mussardo}},\
  and\ \bibinfo {author} {\bibfnamefont {J.~L.}\ \bibnamefont {Cardy}},\
  }\bibfield  {title} {\bibinfo {title} {{The scaling region of the tricritical
  Ising model in two dimensions}},\ }\href
  {https://doi.org/10.1016/0550-3213(91)90206-D} {\bibfield  {journal}
  {\bibinfo  {journal} {Nucl. Phys. B}\ }\textbf {\bibinfo {volume} {348}},\
  \bibinfo {pages} {591} (\bibinfo {year} {1991})}\BibitemShut {NoStop}%
\bibitem [{\citenamefont {Katz}\ \emph {et~al.}(2016)\citenamefont {Katz},
  \citenamefont {Khandker},\ and\ \citenamefont {Walters}}]{Katz:2016hxp}%
  \BibitemOpen
  \bibfield  {author} {\bibinfo {author} {\bibfnamefont {E.}~\bibnamefont
  {Katz}}, \bibinfo {author} {\bibfnamefont {Z.~U.}\ \bibnamefont {Khandker}},\
  and\ \bibinfo {author} {\bibfnamefont {M.~T.}\ \bibnamefont {Walters}},\
  }\bibfield  {title} {\bibinfo {title} {A conformal truncation framework for
  infinite-volume dynamics},\ }\href {https://doi.org/10.1007/JHEP07(2016)140}
  {\bibfield  {journal} {\bibinfo  {journal} {J. High Energy Phys.}\ }\textbf
  {\bibinfo {volume} {2016}}\bibinfo  {number} { (07)},\ \bibinfo {pages}
  {140}}\BibitemShut {NoStop}%
\bibitem [{\citenamefont {Katz}\ \emph
  {et~al.}(2014{\natexlab{a}})\citenamefont {Katz}, \citenamefont
  {Marques~Tavares},\ and\ \citenamefont {Xu}}]{Katz:2014uoa}%
  \BibitemOpen
\bibfield  {number} {  }\bibfield  {author} {\bibinfo {author} {\bibfnamefont
  {E.}~\bibnamefont {Katz}}, \bibinfo {author} {\bibfnamefont {G.}~\bibnamefont
  {Marques~Tavares}},\ and\ \bibinfo {author} {\bibfnamefont {Y.}~\bibnamefont
  {Xu}},\ }\bibfield  {title} {\bibinfo {title} {{A solution of 2D QCD at
  finite $N$ using a conformal basis}},\ }\href
  {https://arxiv.org/abs/1405.6727} {\bibfield  {journal} {\bibinfo  {journal}
  {arXiv:1405.6727}\ } (\bibinfo {year} {2014}{\natexlab{a}})}\BibitemShut
  {NoStop}%
\bibitem [{\citenamefont {Anand}\ \emph
  {et~al.}(2021{\natexlab{a}})\citenamefont {Anand}, \citenamefont
  {Fitzpatrick}, \citenamefont {Katz},\ and\ \citenamefont
  {Xin}}]{Anand:2021qnd}%
  \BibitemOpen
  \bibfield  {author} {\bibinfo {author} {\bibfnamefont {N.}~\bibnamefont
  {Anand}}, \bibinfo {author} {\bibfnamefont {A.~L.}\ \bibnamefont
  {Fitzpatrick}}, \bibinfo {author} {\bibfnamefont {E.}~\bibnamefont {Katz}},\
  and\ \bibinfo {author} {\bibfnamefont {Y.}~\bibnamefont {Xin}},\ }\bibfield
  {title} {\bibinfo {title} {Chiral limit of {2D QCD} revisited with lightcone
  conformal truncation},\ }\href {https://arxiv.org/abs/2111.00021} {\bibfield
  {journal} {\bibinfo  {journal} {arXiv:2111.00021}\ } (\bibinfo {year}
  {2021}{\natexlab{a}})}\BibitemShut {NoStop}%
\bibitem [{\citenamefont {Hogervorst}(2018)}]{Hogervorst:2018otc}%
  \BibitemOpen
  \bibfield  {author} {\bibinfo {author} {\bibfnamefont {M.}~\bibnamefont
  {Hogervorst}},\ }\bibfield  {title} {\bibinfo {title} {{RG flows on $S^d$ and
  Hamiltonian truncation}},\ }\href {https://arxiv.org/abs/1811.00528}
  {\bibfield  {journal} {\bibinfo  {journal} {arXiv:1811.00528}\ } (\bibinfo
  {year} {2018})}\BibitemShut {NoStop}%
\bibitem [{\citenamefont {Hogervorst}\ \emph {et~al.}(2021)\citenamefont
  {Hogervorst}, \citenamefont {Meineri}, \citenamefont {Penedones},\ and\
  \citenamefont {Vaziri}}]{Hogervorst:2021spa}%
  \BibitemOpen
  \bibfield  {author} {\bibinfo {author} {\bibfnamefont {M.}~\bibnamefont
  {Hogervorst}}, \bibinfo {author} {\bibfnamefont {M.}~\bibnamefont {Meineri}},
  \bibinfo {author} {\bibfnamefont {J.}~\bibnamefont {Penedones}},\ and\
  \bibinfo {author} {\bibfnamefont {K.~S.}\ \bibnamefont {Vaziri}},\ }\bibfield
   {title} {\bibinfo {title} {{Hamiltonian truncation in Anti-de Sitter
  spacetime}},\ }\href {https://doi.org/10.1007/JHEP08(2021)063} {\bibfield
  {journal} {\bibinfo  {journal} {J. High Energy Phys.}\ }\textbf {\bibinfo
  {volume} {2021}}\bibinfo  {number} { (08)},\ \bibinfo {pages}
  {063}}\BibitemShut {NoStop}%
\bibitem [{\citenamefont {Chen}\ \emph
  {et~al.}(2022{\natexlab{a}})\citenamefont {Chen}, \citenamefont
  {Fitzpatrick},\ and\ \citenamefont {Karateev}}]{Chen:2021bmm}%
  \BibitemOpen
\bibfield  {number} {  }\bibfield  {author} {\bibinfo {author} {\bibfnamefont
  {H.}~\bibnamefont {Chen}}, \bibinfo {author} {\bibfnamefont {A.~L.}\
  \bibnamefont {Fitzpatrick}},\ and\ \bibinfo {author} {\bibfnamefont
  {D.}~\bibnamefont {Karateev}},\ }\bibfield  {title} {\bibinfo {title} {{Form
  factors and spectral densities from lightcone conformal truncation}},\ }\href
  {https://doi.org/10.1007/JHEP04(2022)109} {\bibfield  {journal} {\bibinfo
  {journal} {J. High Energy Phys.}\ }\textbf {\bibinfo {volume} {2022}}\bibinfo
   {number} { (04)},\ \bibinfo {pages} {109}}\BibitemShut {NoStop}%
\bibitem [{\citenamefont {Katz}\ \emph
  {et~al.}(2014{\natexlab{b}})\citenamefont {Katz}, \citenamefont {Tavares},\
  and\ \citenamefont {Xu}}]{Katz:2013qua}%
  \BibitemOpen
\bibfield  {number} {  }\bibfield  {author} {\bibinfo {author} {\bibfnamefont
  {E.}~\bibnamefont {Katz}}, \bibinfo {author} {\bibfnamefont {G.~M.}\
  \bibnamefont {Tavares}},\ and\ \bibinfo {author} {\bibfnamefont
  {Y.}~\bibnamefont {Xu}},\ }\bibfield  {title} {\bibinfo {title} {{Solving 2D
  QCD with an adjoint fermion analytically}},\ }\href
  {https://doi.org/10.1007/JHEP05(2014)143} {\bibfield  {journal} {\bibinfo
  {journal} {J. High Energy Phys.}\ }\textbf {\bibinfo {volume} {2014}}\bibinfo
   {number} { (5)},\ \bibinfo {pages} {143}}\BibitemShut {NoStop}%
\bibitem [{\citenamefont {Fitzpatrick}\ \emph {et~al.}(2018)\citenamefont
  {Fitzpatrick}, \citenamefont {Kaplan}, \citenamefont {Katz}, \citenamefont
  {Vitale},\ and\ \citenamefont {Walters}}]{Fitzpatrick:2018ttk}%
  \BibitemOpen
\bibfield  {number} {  }\bibfield  {author} {\bibinfo {author} {\bibfnamefont
  {A.~L.}\ \bibnamefont {Fitzpatrick}}, \bibinfo {author} {\bibfnamefont
  {J.}~\bibnamefont {Kaplan}}, \bibinfo {author} {\bibfnamefont
  {E.}~\bibnamefont {Katz}}, \bibinfo {author} {\bibfnamefont {L.~G.}\
  \bibnamefont {Vitale}},\ and\ \bibinfo {author} {\bibfnamefont {M.~T.}\
  \bibnamefont {Walters}},\ }\bibfield  {title} {\bibinfo {title} {{Lightcone
  effective Hamiltonians and RG flows}},\ }\href
  {https://doi.org/10.1007/JHEP08(2018)120} {\bibfield  {journal} {\bibinfo
  {journal} {J. High Energy Phys.}\ }\textbf {\bibinfo {volume} {2018}}\bibinfo
   {number} { (08)},\ \bibinfo {pages} {120}}\BibitemShut {NoStop}%
\bibitem [{\citenamefont {Chen}\ \emph {et~al.}(2025)\citenamefont {Chen},
  \citenamefont {Fitzpatrick}, \citenamefont {Katz},\ and\ \citenamefont
  {Xin}}]{Chen:2022zms}%
  \BibitemOpen
\bibfield  {number} {  }\bibfield  {author} {\bibinfo {author} {\bibfnamefont
  {H.}~\bibnamefont {Chen}}, \bibinfo {author} {\bibfnamefont {A.~L.}\
  \bibnamefont {Fitzpatrick}}, \bibinfo {author} {\bibfnamefont
  {E.}~\bibnamefont {Katz}},\ and\ \bibinfo {author} {\bibfnamefont
  {Y.}~\bibnamefont {Xin}},\ }\bibfield  {title} {\bibinfo {title} {{Giving
  Hamiltonian truncation a boost}},\ }\href
  {https://doi.org/10.1007/JHEP03(2025)043} {\bibfield  {journal} {\bibinfo
  {journal} {J. High Energy Phys.}\ }\textbf {\bibinfo {volume} {2025}}\bibinfo
   {number} { (03)},\ \bibinfo {pages} {043}}\BibitemShut {NoStop}%
\bibitem [{\citenamefont {Fitzpatrick}\ \emph {et~al.}(2025)\citenamefont
  {Fitzpatrick}, \citenamefont {Katz},\ and\ \citenamefont
  {Xin}}]{Fitzpatrick:2023aqm}%
  \BibitemOpen
\bibfield  {number} {  }\bibfield  {author} {\bibinfo {author} {\bibfnamefont
  {A.~L.}\ \bibnamefont {Fitzpatrick}}, \bibinfo {author} {\bibfnamefont
  {E.}~\bibnamefont {Katz}},\ and\ \bibinfo {author} {\bibfnamefont
  {Y.}~\bibnamefont {Xin}},\ }\bibfield  {title} {\bibinfo {title} {{Lightcone
  Hamiltonian for Ising field theory I: $T < T_c$}},\ }\href
  {https://doi.org/10.21468/SciPostPhys.18.6.179} {\bibfield  {journal}
  {\bibinfo  {journal} {SciPost Phys.}\ }\textbf {\bibinfo {volume} {18}},\
  \bibinfo {pages} {179} (\bibinfo {year} {2025})}\BibitemShut {NoStop}%
\bibitem [{\citenamefont {Chen}\ \emph
  {et~al.}(2022{\natexlab{b}})\citenamefont {Chen}, \citenamefont
  {Fitzpatrick},\ and\ \citenamefont {Karateev}}]{Chen:2021pgx}%
  \BibitemOpen
  \bibfield  {author} {\bibinfo {author} {\bibfnamefont {H.}~\bibnamefont
  {Chen}}, \bibinfo {author} {\bibfnamefont {A.~L.}\ \bibnamefont
  {Fitzpatrick}},\ and\ \bibinfo {author} {\bibfnamefont {D.}~\bibnamefont
  {Karateev}},\ }\bibfield  {title} {\bibinfo {title} {{Bootstrapping 2D
  {\ensuremath{\phi}}$^{4}$ theory with Hamiltonian truncation data}},\ }\href
  {https://doi.org/10.1007/JHEP02(2022)146} {\bibfield  {journal} {\bibinfo
  {journal} {J. High Energy Phys.}\ }\textbf {\bibinfo {volume} {2022}}\bibinfo
   {number} { (02)},\ \bibinfo {pages} {146}}\BibitemShut {NoStop}%
\bibitem [{\citenamefont {Elias-Mir\'o}\ and\ \citenamefont
  {Hardy}(2020)}]{PhysRevD.102.065001}%
  \BibitemOpen
\bibfield  {number} {  }\bibfield  {author} {\bibinfo {author} {\bibfnamefont
  {J.}~\bibnamefont {Elias-Mir\'o}}\ and\ \bibinfo {author} {\bibfnamefont
  {E.}~\bibnamefont {Hardy}},\ }\bibfield  {title} {\bibinfo {title}
  {{Exploring Hamiltonian truncation in $d=2+1$}},\ }\href
  {https://doi.org/10.1103/PhysRevD.102.065001} {\bibfield  {journal} {\bibinfo
   {journal} {Phys. Rev. D}\ }\textbf {\bibinfo {volume} {102}},\ \bibinfo
  {pages} {065001} (\bibinfo {year} {2020})}\BibitemShut {NoStop}%
\bibitem [{\citenamefont {Anand}\ \emph
  {et~al.}(2021{\natexlab{b}})\citenamefont {Anand}, \citenamefont {Katz},
  \citenamefont {Khandker},\ and\ \citenamefont {Walters}}]{Anand:2020qnp}%
  \BibitemOpen
  \bibfield  {author} {\bibinfo {author} {\bibfnamefont {N.}~\bibnamefont
  {Anand}}, \bibinfo {author} {\bibfnamefont {E.}~\bibnamefont {Katz}},
  \bibinfo {author} {\bibfnamefont {Z.~U.}\ \bibnamefont {Khandker}},\ and\
  \bibinfo {author} {\bibfnamefont {M.~T.}\ \bibnamefont {Walters}},\
  }\bibfield  {title} {\bibinfo {title} {{Nonperturbative dynamics of (2+1)D
  $\phi^4$-theory from Hamiltonian truncation}},\ }\href
  {https://doi.org/10.1007/JHEP05(2021)190} {\bibfield  {journal} {\bibinfo
  {journal} {J. High Energy Phys.}\ }\textbf {\bibinfo {volume} {2021}}\bibinfo
   {number} { (5)},\ \bibinfo {pages} {190}}\BibitemShut {NoStop}%
\bibitem [{\citenamefont {Rakovszky}\ \emph {et~al.}(2016)\citenamefont
  {Rakovszky}, \citenamefont {Mesty{\'a}n}, \citenamefont {Collura},
  \citenamefont {Kormos},\ and\ \citenamefont
  {Tak{\'a}cs}}]{Rakovszky:2016ugs}%
  \BibitemOpen
\bibfield  {number} {  }\bibfield  {author} {\bibinfo {author} {\bibfnamefont
  {T.}~\bibnamefont {Rakovszky}}, \bibinfo {author} {\bibfnamefont
  {M.}~\bibnamefont {Mesty{\'a}n}}, \bibinfo {author} {\bibfnamefont
  {M.}~\bibnamefont {Collura}}, \bibinfo {author} {\bibfnamefont
  {M.}~\bibnamefont {Kormos}},\ and\ \bibinfo {author} {\bibfnamefont
  {G.}~\bibnamefont {Tak{\'a}cs}},\ }\bibfield  {title} {\bibinfo {title}
  {{Hamiltonian truncation approach to quenches in the Ising field theory}},\
  }\href {https://doi.org/10.1016/j.nuclphysb.2016.08.024} {\bibfield
  {journal} {\bibinfo  {journal} {Nucl. Phys. B}\ }\textbf {\bibinfo {volume}
  {911}},\ \bibinfo {pages} {805} (\bibinfo {year} {2016})}\BibitemShut
  {NoStop}%
\bibitem [{\citenamefont {H{\'o}ds{\'a}gi}\ \emph {et~al.}(2018)\citenamefont
  {H{\'o}ds{\'a}gi}, \citenamefont {Kormos},\ and\ \citenamefont
  {Tak{\'a}cs}}]{Hodsagi:2018sul}%
  \BibitemOpen
  \bibfield  {author} {\bibinfo {author} {\bibfnamefont {K.}~\bibnamefont
  {H{\'o}ds{\'a}gi}}, \bibinfo {author} {\bibfnamefont {M.}~\bibnamefont
  {Kormos}},\ and\ \bibinfo {author} {\bibfnamefont {G.}~\bibnamefont
  {Tak{\'a}cs}},\ }\bibfield  {title} {\bibinfo {title} {{Quench dynamics of
  the Ising field theory in a magnetic field}},\ }\href
  {https://doi.org/10.21468/SciPostPhys.5.3.027} {\bibfield  {journal}
  {\bibinfo  {journal} {SciPost Phys.}\ }\textbf {\bibinfo {volume} {5}},\
  \bibinfo {pages} {027} (\bibinfo {year} {2018})}\BibitemShut {NoStop}%
\bibitem [{\citenamefont {Konik}\ and\ \citenamefont
  {Adamov}(2007)}]{PhysRevLett.98.147205}%
  \BibitemOpen
  \bibfield  {author} {\bibinfo {author} {\bibfnamefont {R.~M.}\ \bibnamefont
  {Konik}}\ and\ \bibinfo {author} {\bibfnamefont {Y.}~\bibnamefont {Adamov}},\
  }\bibfield  {title} {\bibinfo {title} {Numerical renormalization group for
  continuum one-dimensional systems},\ }\href
  {https://doi.org/10.1103/PhysRevLett.98.147205} {\bibfield  {journal}
  {\bibinfo  {journal} {Phys. Rev. Lett.}\ }\textbf {\bibinfo {volume} {98}},\
  \bibinfo {pages} {147205} (\bibinfo {year} {2007})}\BibitemShut {NoStop}%
\bibitem [{\citenamefont {Brandino}\ \emph {et~al.}(2010)\citenamefont
  {Brandino}, \citenamefont {Konik},\ and\ \citenamefont
  {Mussardo}}]{Brandino:2010sv}%
  \BibitemOpen
  \bibfield  {author} {\bibinfo {author} {\bibfnamefont {G.~P.}\ \bibnamefont
  {Brandino}}, \bibinfo {author} {\bibfnamefont {R.~M.}\ \bibnamefont
  {Konik}},\ and\ \bibinfo {author} {\bibfnamefont {G.}~\bibnamefont
  {Mussardo}},\ }\bibfield  {title} {\bibinfo {title} {{Energy level
  distribution of perturbed conformal field theories}},\ }\href
  {https://doi.org/10.1088/1742-5468/2010/07/P07013} {\bibfield  {journal}
  {\bibinfo  {journal} {J. Stat. Mech.}\ }\textbf {\bibinfo {volume} {1007}},\
  \bibinfo {pages} {P07013} (\bibinfo {year} {2010})}\BibitemShut {NoStop}%
\bibitem [{\citenamefont {Albert}\ \emph {et~al.}(2024)\citenamefont {Albert},
  \citenamefont {Zhang},\ and\ \citenamefont {Tu}}]{Albert:2024pwv}%
  \BibitemOpen
  \bibfield  {author} {\bibinfo {author} {\bibfnamefont {N.}~\bibnamefont
  {Albert}}, \bibinfo {author} {\bibfnamefont {Y.}~\bibnamefont {Zhang}},\ and\
  \bibinfo {author} {\bibfnamefont {H.-H.}\ \bibnamefont {Tu}},\ }\bibfield
  {title} {\bibinfo {title} {{Truncated Gaussian basis approach for simulating
  many-body dynamics}},\ }\href {https://arxiv.org/abs/2410.04204} {\bibfield
  {journal} {\bibinfo  {journal} {arXiv:2410.04204}\ } (\bibinfo {year}
  {2024})}\BibitemShut {NoStop}%
\bibitem [{\citenamefont {Zhu}\ \emph {et~al.}(2023)\citenamefont {Zhu},
  \citenamefont {Han}, \citenamefont {Huffman}, \citenamefont {Hofmann},\ and\
  \citenamefont {He}}]{PhysRevX.13.021009}%
  \BibitemOpen
  \bibfield  {author} {\bibinfo {author} {\bibfnamefont {W.}~\bibnamefont
  {Zhu}}, \bibinfo {author} {\bibfnamefont {C.}~\bibnamefont {Han}}, \bibinfo
  {author} {\bibfnamefont {E.}~\bibnamefont {Huffman}}, \bibinfo {author}
  {\bibfnamefont {J.~S.}\ \bibnamefont {Hofmann}},\ and\ \bibinfo {author}
  {\bibfnamefont {Y.-C.}\ \bibnamefont {He}},\ }\bibfield  {title} {\bibinfo
  {title} {{Uncovering conformal symmetry in the 3D Ising transition:
  State-operator correspondence from a quantum fuzzy sphere regularization}},\
  }\href {https://doi.org/10.1103/PhysRevX.13.021009} {\bibfield  {journal}
  {\bibinfo  {journal} {Phys. Rev. X}\ }\textbf {\bibinfo {volume} {13}},\
  \bibinfo {pages} {021009} (\bibinfo {year} {2023})}\BibitemShut {NoStop}%
\bibitem [{\citenamefont {Hu}\ \emph {et~al.}(2024)\citenamefont {Hu},
  \citenamefont {He},\ and\ \citenamefont {Zhu}}]{Hu2024}%
  \BibitemOpen
  \bibfield  {author} {\bibinfo {author} {\bibfnamefont {L.}~\bibnamefont
  {Hu}}, \bibinfo {author} {\bibfnamefont {Y.-C.}\ \bibnamefont {He}},\ and\
  \bibinfo {author} {\bibfnamefont {W.}~\bibnamefont {Zhu}},\ }\bibfield
  {title} {\bibinfo {title} {{Solving conformal defects in 3D conformal field
  theory using fuzzy sphere regularization}},\ }\href
  {https://doi.org/10.1038/s41467-024-47978-y} {\bibfield  {journal} {\bibinfo
  {journal} {Nat. Commun.}\ }\textbf {\bibinfo {volume} {15}},\ \bibinfo
  {pages} {3659} (\bibinfo {year} {2024})}\BibitemShut {NoStop}%
\bibitem [{\citenamefont {Yang}\ \emph {et~al.}(2023)\citenamefont {Yang},
  \citenamefont {Xie}, \citenamefont {Nikitin}, \citenamefont {Wu},\ and\
  \citenamefont {Podlesnyak}}]{PhysRevB.108.L020402}%
  \BibitemOpen
  \bibfield  {author} {\bibinfo {author} {\bibfnamefont {J.}~\bibnamefont
  {Yang}}, \bibinfo {author} {\bibfnamefont {T.}~\bibnamefont {Xie}}, \bibinfo
  {author} {\bibfnamefont {S.~E.}\ \bibnamefont {Nikitin}}, \bibinfo {author}
  {\bibfnamefont {J.}~\bibnamefont {Wu}},\ and\ \bibinfo {author}
  {\bibfnamefont {A.}~\bibnamefont {Podlesnyak}},\ }\bibfield  {title}
  {\bibinfo {title} {{Confinement of many-body Bethe strings}},\ }\href
  {https://doi.org/10.1103/PhysRevB.108.L020402} {\bibfield  {journal}
  {\bibinfo  {journal} {Phys. Rev. B}\ }\textbf {\bibinfo {volume} {108}},\
  \bibinfo {pages} {L020402} (\bibinfo {year} {2023})}\BibitemShut {NoStop}%
\bibitem [{\citenamefont {Yang}\ and\ \citenamefont
  {Wu}(2024)}]{PhysRevB.109.214421}%
  \BibitemOpen
  \bibfield  {author} {\bibinfo {author} {\bibfnamefont {J.}~\bibnamefont
  {Yang}}\ and\ \bibinfo {author} {\bibfnamefont {J.}~\bibnamefont {Wu}},\
  }\bibfield  {title} {\bibinfo {title} {{Truncated string state space approach
  and its application to the nonintegrable spin-1/2 Heisenberg chain}},\ }\href
  {https://doi.org/10.1103/PhysRevB.109.214421} {\bibfield  {journal} {\bibinfo
   {journal} {Phys. Rev. B}\ }\textbf {\bibinfo {volume} {109}},\ \bibinfo
  {pages} {214421} (\bibinfo {year} {2024})}\BibitemShut {NoStop}%
\bibitem [{\citenamefont {Wang}\ \emph {et~al.}(2025)\citenamefont {Wang},
  \citenamefont {He},\ and\ \citenamefont {Wu}}]{rw2m-33v6}%
  \BibitemOpen
  \bibfield  {author} {\bibinfo {author} {\bibfnamefont {X.}~\bibnamefont
  {Wang}}, \bibinfo {author} {\bibfnamefont {X.}~\bibnamefont {He}},\ and\
  \bibinfo {author} {\bibfnamefont {J.}~\bibnamefont {Wu}},\ }\bibfield
  {title} {\bibinfo {title} {Berry connection and quantum geometry in
  time-dependent systems with instantaneous quantum integrable field theory},\
  }\href {https://doi.org/10.1103/rw2m-33v6} {\bibfield  {journal} {\bibinfo
  {journal} {Phys. Rev. B}\ }\textbf {\bibinfo {volume} {112}},\ \bibinfo
  {pages} {L201102} (\bibinfo {year} {2025})}\BibitemShut {NoStop}%
\bibitem [{\citenamefont {Sachdev}(2011)}]{sachdev_2011}%
  \BibitemOpen
  \bibfield  {author} {\bibinfo {author} {\bibfnamefont {S.}~\bibnamefont
  {Sachdev}},\ }\href {https://doi.org/10.1017/CBO9780511973765} {\emph
  {\bibinfo {title} {Quantum Phase Transitions}}}\ (\bibinfo  {publisher}
  {Cambridge University Press},\ \bibinfo {address} {Cambridge, England},\
  \bibinfo {year} {2011})\BibitemShut {NoStop}%
\bibitem [{\citenamefont {Pfeuty}(1970)}]{Pfeuty}%
  \BibitemOpen
  \bibfield  {author} {\bibinfo {author} {\bibfnamefont {P.}~\bibnamefont
  {Pfeuty}},\ }\bibfield  {title} {\bibinfo {title} {{The one-dimensional Ising
  model with a transverse field}},\ }\href
  {https://doi.org/10.1016/0003-4916(70)90270-8} {\bibfield  {journal}
  {\bibinfo  {journal} {Ann. Phys.}\ }\textbf {\bibinfo {volume} {59}},\
  \bibinfo {pages} {79} (\bibinfo {year} {1970})}\BibitemShut {NoStop}%
\bibitem [{\citenamefont {Iorgov}\ \emph {et~al.}(2011)\citenamefont {Iorgov},
  \citenamefont {Shadura},\ and\ \citenamefont {Tykhyy}}]{Iorgov_2011}%
  \BibitemOpen
  \bibfield  {author} {\bibinfo {author} {\bibfnamefont {N.}~\bibnamefont
  {Iorgov}}, \bibinfo {author} {\bibfnamefont {V.}~\bibnamefont {Shadura}},\
  and\ \bibinfo {author} {\bibfnamefont {Y.}~\bibnamefont {Tykhyy}},\
  }\bibfield  {title} {\bibinfo {title} {{Spin operator matrix elements in the
  quantum Ising chain: Fermion approach}},\ }\href
  {https://doi.org/10.1088/1742-5468/2011/02/P02028} {\bibfield  {journal}
  {\bibinfo  {journal} {J. Stat. Mech.}\ }\textbf {\bibinfo {volume} {1102}},\
  \bibinfo {pages} {P02028} (\bibinfo {year} {2011})}\BibitemShut {NoStop}%
\bibitem [{\citenamefont
  {Zamolodchikov}(1989{\natexlab{a}})}]{Zamolodchikov:1989fp}%
  \BibitemOpen
  \bibfield  {author} {\bibinfo {author} {\bibfnamefont {A.~B.}\ \bibnamefont
  {Zamolodchikov}},\ }\bibfield  {title} {\bibinfo {title} {{Integrals of
  motion and S-matrix of the (scaled) $T=T_{c}$ Ising model with magnetic
  field}},\ }\href {https://doi.org/10.1142/S0217751X8900176X} {\bibfield
  {journal} {\bibinfo  {journal} {Int. J. Mod. Phys. A}\ }\textbf {\bibinfo
  {volume} {4}},\ \bibinfo {pages} {4235} (\bibinfo {year}
  {1989}{\natexlab{a}})}\BibitemShut {NoStop}%
\bibitem [{\citenamefont {Kj\"all}\ \emph {et~al.}(2011)\citenamefont
  {Kj\"all}, \citenamefont {Pollmann},\ and\ \citenamefont
  {Moore}}]{PhysRevB.83.020407}%
  \BibitemOpen
  \bibfield  {author} {\bibinfo {author} {\bibfnamefont {J.~A.}\ \bibnamefont
  {Kj\"all}}, \bibinfo {author} {\bibfnamefont {F.}~\bibnamefont {Pollmann}},\
  and\ \bibinfo {author} {\bibfnamefont {J.~E.}\ \bibnamefont {Moore}},\
  }\bibfield  {title} {\bibinfo {title} {{Bound states and ${E}_{8}$ symmetry
  effects in perturbed quantum Ising chains}},\ }\href
  {https://doi.org/10.1103/PhysRevB.83.020407} {\bibfield  {journal} {\bibinfo
  {journal} {Phys. Rev. B}\ }\textbf {\bibinfo {volume} {83}},\ \bibinfo
  {pages} {020407} (\bibinfo {year} {2011})}\BibitemShut {NoStop}%
\bibitem [{\citenamefont
  {Zamolodchikov}(1989{\natexlab{b}})}]{Zamolodchikov:1989hfa}%
  \BibitemOpen
  \bibfield  {author} {\bibinfo {author} {\bibfnamefont {A.~B.}\ \bibnamefont
  {Zamolodchikov}},\ }\bibfield  {title} {\bibinfo {title} {{Integrable field
  theory from conformal field theory}},\ }\href
  {https://projecteuclid.org/ebooks/advanced-studies-in-pure-mathematics/Integrable-Systems-in-Quantum-Field-Theory-and-Statistical-Mechanics/chapter/Integrable-Field-Theory-from-Conformal-Field-Theory/10.2969/aspm/01910641}
  {\bibfield  {journal} {\bibinfo  {journal} {Adv. Stud. Pure Math.}\ }\textbf
  {\bibinfo {volume} {19}},\ \bibinfo {pages} {641} (\bibinfo {year}
  {1989}{\natexlab{b}})}\BibitemShut {NoStop}%
\bibitem [{\citenamefont {Wang}\ \emph {et~al.}(2021)\citenamefont {Wang},
  \citenamefont {Zou}, \citenamefont {H\'ods\'agi}, \citenamefont {Kormos},
  \citenamefont {Tak\'acs},\ and\ \citenamefont {Wu}}]{PhysRevB.103.235117}%
  \BibitemOpen
  \bibfield  {author} {\bibinfo {author} {\bibfnamefont {X.}~\bibnamefont
  {Wang}}, \bibinfo {author} {\bibfnamefont {H.}~\bibnamefont {Zou}}, \bibinfo
  {author} {\bibfnamefont {K.}~\bibnamefont {H\'ods\'agi}}, \bibinfo {author}
  {\bibfnamefont {M.}~\bibnamefont {Kormos}}, \bibinfo {author} {\bibfnamefont
  {G.}~\bibnamefont {Tak\'acs}},\ and\ \bibinfo {author} {\bibfnamefont
  {J.}~\bibnamefont {Wu}},\ }\bibfield  {title} {\bibinfo {title} {{Cascade of
  singularities in the spin dynamics of a perturbed quantum critical Ising
  chain}},\ }\href {https://doi.org/10.1103/PhysRevB.103.235117} {\bibfield
  {journal} {\bibinfo  {journal} {Phys. Rev. B}\ }\textbf {\bibinfo {volume}
  {103}},\ \bibinfo {pages} {235117} (\bibinfo {year} {2021})}\BibitemShut
  {NoStop}%
\bibitem [{\citenamefont {LeClair}\ \emph {et~al.}(1998)\citenamefont
  {LeClair}, \citenamefont {Ludwig},\ and\ \citenamefont
  {Mussardo}}]{LeClair:1997gv}%
  \BibitemOpen
  \bibfield  {author} {\bibinfo {author} {\bibfnamefont {A.}~\bibnamefont
  {LeClair}}, \bibinfo {author} {\bibfnamefont {A.}~\bibnamefont {Ludwig}},\
  and\ \bibinfo {author} {\bibfnamefont {G.}~\bibnamefont {Mussardo}},\
  }\bibfield  {title} {\bibinfo {title} {{Integrability of coupled conformal
  field theories}},\ }\href {https://doi.org/10.1016/S0550-3213(97)00724-4}
  {\bibfield  {journal} {\bibinfo  {journal} {Nucl. Phys. B}\ }\textbf
  {\bibinfo {volume} {512}},\ \bibinfo {pages} {523} (\bibinfo {year}
  {1998})}\BibitemShut {NoStop}%
\bibitem [{\citenamefont {Gao}\ \emph {et~al.}(2025{\natexlab{a}})\citenamefont
  {Gao}, \citenamefont {Wang}, \citenamefont {Xi}, \citenamefont {Jiang},
  \citenamefont {Yu},\ and\ \citenamefont {Wu}}]{Gao:2024ngb}%
  \BibitemOpen
  \bibfield  {author} {\bibinfo {author} {\bibfnamefont {Y.}~\bibnamefont
  {Gao}}, \bibinfo {author} {\bibfnamefont {X.}~\bibnamefont {Wang}}, \bibinfo
  {author} {\bibfnamefont {N.}~\bibnamefont {Xi}}, \bibinfo {author}
  {\bibfnamefont {Y.}~\bibnamefont {Jiang}}, \bibinfo {author} {\bibfnamefont
  {R.}~\bibnamefont {Yu}},\ and\ \bibinfo {author} {\bibfnamefont
  {J.}~\bibnamefont {Wu}},\ }\bibfield  {title} {\bibinfo {title} {{Spin
  dynamics and dark particle in a weak-coupled quantum Ising ladder with
  ${\mathcal{D}}_{8}^{(1)}$ spectrum}},\ }\href
  {https://doi.org/10.1103/PhysRevB.111.L201117} {\bibfield  {journal}
  {\bibinfo  {journal} {Phys. Rev. B}\ }\textbf {\bibinfo {volume} {111}},\
  \bibinfo {pages} {L201117} (\bibinfo {year}
  {2025}{\natexlab{a}})}\BibitemShut {NoStop}%
\bibitem [{\citenamefont {Gao}\ \emph {et~al.}(2025{\natexlab{b}})\citenamefont
  {Gao}, \citenamefont {Jiang},\ and\ \citenamefont {Wu}}]{Gao:2025mcg}%
  \BibitemOpen
  \bibfield  {author} {\bibinfo {author} {\bibfnamefont {Y.}~\bibnamefont
  {Gao}}, \bibinfo {author} {\bibfnamefont {Y.}~\bibnamefont {Jiang}},\ and\
  \bibinfo {author} {\bibfnamefont {J.}~\bibnamefont {Wu}},\ }\bibfield
  {title} {\bibinfo {title} {{Mesons in a quantum Ising ladder}},\ }\href
  {https://doi.org/10.1007/JHEP07(2025)072} {\bibfield  {journal} {\bibinfo
  {journal} {J. High Energy Phys.}\ }\textbf {\bibinfo {volume} {2025}}\bibinfo
   {number} { (07)},\ \bibinfo {pages} {072}}\BibitemShut {NoStop}%
\bibitem [{\citenamefont {Takhtadzhan}\ and\ \citenamefont
  {Faddeev}(1979)}]{Takhtajan:1979iv}%
  \BibitemOpen
\bibfield  {number} {  }\bibfield  {author} {\bibinfo {author} {\bibfnamefont
  {L.~A.}\ \bibnamefont {Takhtadzhan}}\ and\ \bibinfo {author} {\bibfnamefont
  {L.~D.}\ \bibnamefont {Faddeev}},\ }\bibfield  {title} {\bibinfo {title}
  {{The quantum method of the inverse problem and the Heisenberg XYZ model}},\
  }\href {https://doi.org/10.1070/RM1979v034n05ABEH003909} {\bibfield
  {journal} {\bibinfo  {journal} {Russ. Math. Surveys}\ }\textbf {\bibinfo
  {volume} {34}},\ \bibinfo {pages} {11} (\bibinfo {year} {1979})}\BibitemShut
  {NoStop}%
\bibitem [{\citenamefont {Yang}(1967)}]{PhysRevLett.19.1312}%
  \BibitemOpen
  \bibfield  {author} {\bibinfo {author} {\bibfnamefont {C.~N.}\ \bibnamefont
  {Yang}},\ }\bibfield  {title} {\bibinfo {title} {Some exact results for the
  many-body problem in one dimension with repulsive delta-function
  interaction},\ }\href {https://doi.org/10.1103/PhysRevLett.19.1312}
  {\bibfield  {journal} {\bibinfo  {journal} {Phys. Rev. Lett.}\ }\textbf
  {\bibinfo {volume} {19}},\ \bibinfo {pages} {1312} (\bibinfo {year}
  {1967})}\BibitemShut {NoStop}%
\bibitem [{\citenamefont {Kitanine}\ \emph {et~al.}(1999)\citenamefont
  {Kitanine}, \citenamefont {Maillet},\ and\ \citenamefont
  {Terras}}]{KITANINE1999647}%
  \BibitemOpen
  \bibfield  {author} {\bibinfo {author} {\bibfnamefont {N.}~\bibnamefont
  {Kitanine}}, \bibinfo {author} {\bibfnamefont {J.}~\bibnamefont {Maillet}},\
  and\ \bibinfo {author} {\bibfnamefont {V.}~\bibnamefont {Terras}},\
  }\bibfield  {title} {\bibinfo {title} {{Form factors of the XXZ Heisenberg
  spin-1/2 finite chain}},\ }\href
  {https://doi.org/https://doi.org/10.1016/S0550-3213(99)00295-3} {\bibfield
  {journal} {\bibinfo  {journal} {Nucl. Phys. B}\ }\textbf {\bibinfo {volume}
  {554}},\ \bibinfo {pages} {647} (\bibinfo {year} {1999})}\BibitemShut
  {NoStop}%
\bibitem [{\citenamefont {Slavnov}(1990)}]{slavnov_1990}%
  \BibitemOpen
  \bibfield  {author} {\bibinfo {author} {\bibfnamefont {N.}~\bibnamefont
  {Slavnov}},\ }\bibfield  {title} {\bibinfo {title} {{Nonequal-time current
  correlation function in~a~one-dimensional {B}ose gas}},\ }\href
  {https://doi.org/10.1007/BF01029221} {\bibfield  {journal} {\bibinfo
  {journal} {Teor. Mat. Fiz.}\ }\textbf {\bibinfo {volume} {82}},\ \bibinfo
  {pages} {389} (\bibinfo {year} {1990})}\BibitemShut {NoStop}%
\end{thebibliography}%

\end{document}